\documentclass[12pt,aps,prc,showpacs,nofootinbib]{revtex4}%
\usepackage{amssymb}
\usepackage{makeidx}
\usepackage{amsmath}
\usepackage{amsfonts}
\usepackage{graphicx}
\usepackage[T2A]{fontenc}
\usepackage[cp1251]{inputenc}

\begin{document}

\title{The contribution of pseudoscalar mesons\\ 
to hyperfine structure of muonic hydrogen}
\author{A.~E.~Dorokhov\footnote{E-mail:~dorokhov@theor.jinr.ru}}
\affiliation{Joint Institute of Nuclear Research, BLTP,\\
141980, Moscow region, Dubna, Russia}
\author{N.~I.~Kochelev\footnote{E-mail:~nikkochelev@mail.ru}}
\affiliation{Institute of Modern Physics of Chinese Academy of Sciences, 730000, Lanzhou, China}
\affiliation{Joint Institute of Nuclear Research, BLTP,\\
141980, Moscow region, Dubna, Russia}
\author{A.~P.~Martynenko\footnote{E-mail:~a.p.martynenko@samsu.ru}}
\affiliation{Samara University, 443086, Samara, Russia}
\author{F.~A.~Martynenko}
\affiliation{Samara University, 443086, Samara, Russia}
\author{R.~N.~Faustov}
\affiliation{Institute of Informatics in Education, FRC CSC RAS, 119333, Moscow, Russia}

\begin{abstract}
In the framework of the quasipotential method in quantum electrodynamics
we calculate the contribution of pseudoscalar mesons to the interaction operator
of a muon and a proton in muonic hydrogen atom.
The parametrization of the transition form factor of two photons into
$ \pi $, $ \eta $ mesons, based on the experimental data on the transition form factors 
and QCD asymptotics is used.
Numerical estimates of the contributions to the hyperfine
structure of the spectrum of the S and P levels are presented.
\end{abstract}

\pacs{31.30.Jv, 12.20.Ds, 32.10.Fn}

\maketitle

\section{Introduction}

Precise investigation of the Lamb shift and hyperfine structure
of light muonic atoms is a fundamental problem for testing the Standard model and establishing 
the exact values of its parameters, as well as searching for effects of new physics. 
At present, the relevance of these studies is primarily related to
experiments conducted by the collaboration
CREMA (Charge Radius Experiments with Muonic Atoms) 
\cite{crema1,crema2,crema3,crema4} with muonic hydrogen and deuterium by methods
of laser spectroscopy.
So, as a result of measuring the transition frequency $ 2P^{F=2}_{3/2}-2S^{F=1}_{1/2}$
a more accurate value of the proton charge radius was found to be $r_p = 0.84184(67)$ fm,
which is different from the value recommended by CODATA for $7\sigma$ 
\cite{Mohr:2012tt}. The CODATA value is based
on the spectroscopy of the electronic hydrogen atom and on electron-nucleon
scattering. The measurement of the transition frequency 
$2P^{F=1}_{3/2}-2S^{F=0}_{1/2}$ for the singlet
$2S$ of the state $ (\mu p)$ allowed to obtain the hyperfine splitting of the $2S$ energy level
in muonic hydrogen, and also the values of the Zemach's radius $r_Z=1.082(37)$ fm and magnetic
radius $r_M=0.87(6)$ fm. The first measurement of three transition 
frequencies between energy levels $2P$ and $2S $ for
muonic deuterium $(2S_{1/2}^{F=3/2}-2P_{3/2}^{F=5/2})$, $(2S_{1/2}^{F =1/2}-2P_{3/2}^{F =3/2})$, 
$ (2S_{1/2}^{F=1/2}-2P_{3/2}^{F=1/2})$
allowed us to obtain in 2.7 times the more accurate value of the charge radius of the deuteron,
which is also less than the value recommended by CODATA \cite{Mohr:2012tt}, by $7.5 \sigma$ 
\cite{crema4}. 
As a result, a situation emerges when there is an inexplicable discrepancy between the values 
of such fundamental
parameters, like the charge radius of a proton and deuteron, obtained from electronic and muonic
atoms. In the process of searching for possible solutions of the proton charge radius "puzzle"
various hypotheses were formulated, including the idea of the nonuniversality of the interaction 
of electrons and muons with nucleons. It is possible that the inclusion in experimental
studies of such muonic atoms as muonic helium $ (\mu^3_2He)^+ $, muonic tritium
$(\mu t)$ with nuclei consisting of three nucleons, or other light muonic
atoms will clarify the problem.
In the experiments of the CREMA collaboration one very important task is posed: 
to obtain an order of magnitude more accurate values of the charge radii of the 
simplest nuclei (proton, deuteron, helion, alpha particle ....) that enter 
into one form or another into theoretical expressions for intervals of fine 
or hyperfine structure of the spectrum.
In this case, the high sensitivity of the characteristics of the bound muon to
distribution of charge density and magnetic moment of the nucleus is used.
Successful realization of this program is possible only in combination with precise theoretical
calculations of various energy intervals, measured experimentally. In this way,
the problem of a more accurate theoretical construction of the particle interaction operator in
quantum electrodynamics, the calculation of new corrections in the energy spectrum of muonic atoms
acquires a special urgency.

\section{General formalism}

To study the fine and hyperfine structure of the spectrum of the muonic hydrogen, 
we use a quasipotential method in quantum electrodynamics in which the bound state 
of a muon and a proton is described in the leading order in the fine-structure constant 
by the Schr\"odinger equation with the Coulomb potential \cite{apm2005,apm1999,pra2016}. 
The first part of the important corrections in the energy spectrum of the S- and P-states 
is determined by the Breit Hamiltonian \cite{apm2005,apm1999,t4} 
(further, the abbreviation "fs" and "hfs" is used to denote the contribution to the fine 
structure and hyperfine structure of the energy spectrum):
\begin{equation}
H_B=H_0+\Delta V_B^{fs}+\Delta V_B^{hfs},~~~H_0=\frac{{\bf p}^2}{2\mu}-\frac{Z\alpha}{r},
\label{eq:1}
\end{equation}
\begin{equation}
\label{eq:2}
\Delta V_B^{fs}=-\frac{{\bf p}^4}{8m_l^3}-\frac{{\bf p}^4}{8m_p^3}+\frac{\pi Z\alpha}{2}
\left(\frac{1}{m_l^2}+\frac{1}{m_p^2}\right)\delta({\bf r})-\frac{Z\alpha}{2m_lm_pr}
\left({\bf p}^2+\frac{{\bf r}({\bf r}{\bf p}){\bf p}}{r^2}\right)+
\end{equation}
\begin{displaymath}
+\frac{Z\alpha}{2m_l^2r^3}\left[1+\frac{2m_l}{m_p}+2a_\mu\left(1+\frac{m_l}{m_p}\right)\right]
({\bf L}{\bf s}_1),
\end{displaymath}
\begin{equation}
\label{eq:3}
\Delta V_B^{hfs}=\frac{8\pi\alpha\mu_p}{3m_lm_p}({\bf s}_1{\bf s}_2)\delta({\bf r})+
-\frac{\alpha\mu_p(1+a_\mu)}{m_lm_pr^3}\left[({\bf s}_1{\bf s}_2)-3({\bf s}_1{\bf n})
({\bf s}_2{\bf n})\right]+
\end{equation}
\begin{displaymath}
\frac{\alpha\mu_p}{m_lm_pr^3}\left[1+\frac{m_l}{m_p}-\frac{m_l}{2m_p\mu_p}\right]({\bf L}{\bf s}_2)
\end{displaymath}
where $m_l$, $m_p$ are the masses of muon and proton correspondingly, 
$\mu_p$ is the proton magnetic moment,
${\bf s}_1$ и ${\bf s}_2$ are the muon and proton spins.
The contribution of interactions (\ref{eq:1})-(\ref{eq:3}) to the energy spectrum 
of different muonic atoms is well studied \cite{egs,borie3,kp1,uj,kkis,apmjetp,apm2008,apm2011}. 
The operator (\ref{eq:3}) gives the main contribution of the order $\alpha^4 $ 
to the hyperfine structure of the energy spectrum of the muonic atom (Fermi energy). 
The precise calculation of the hyperfine structure, which is necessary for comparison 
with the experimental data, requires the consideration of various corrections.

\begin{figure}[th]
\centerline{\includegraphics[scale=1.]{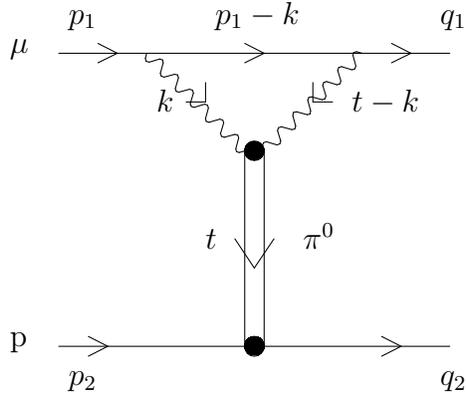}}
\caption{The amplitude of $\pi$, $\eta$, $\eta'$ interaction in muonic hydrogen.}%
\label{fig1}%
\end{figure}

An infinite series of perturbation theory for the particle interaction 
operator contains the contributions of different interactions. One such 
contribution due to the exchange of a pseudoscalar meson is investigated 
in this paper. The amplitude of this interaction is shown in Fig.~\ref{fig1}.

The effective vertex of the interaction of the $\pi^0$ meson 
(or other pseudoscalar mesons $\eta$, $\eta'$) and virtual photons 
can be expressed in terms of the transition form factor 
$F_{\pi^0\gamma^\ast\gamma^\ast}(k_1^2,k_2^2)$ in the form:
\begin{equation}
\label{eq:4}
V^{\mu\nu}(k_1,k_2)=i\varepsilon^{\mu\nu\alpha\beta}k_{1\alpha}k_{2\beta}
\frac{\alpha}{\pi F_\pi}F_{\pi^0\gamma^\ast\gamma^\ast}(k_1^2,k_2^2),
\end{equation}
where $k_1$, $k_2$ are four-momenta of virtual photons. 
The transition form factor is normalized by the condition: 
$F_{\pi^0\gamma^\ast\gamma^\ast}(0,0)=1$.
With increasing $k_1^2 $, $k_2^2$, the function rapidly 
decreases, which ensures the ultraviolet convergence of the loop integral 
in the interaction amplitude. The contribution of pseudoscalar mesons 
to hadronic light-by-light scattering was studied earlier in 
the calculation of the anomalous magnetic moment of the muon 
and the hyperfine structure of muonium 
\cite{nyffeler,Dorokhov:2008pw,dorokhov1,dorokhov3,Dorokhov:2009xs,fm2002,sgk2008}.

Let us first consider the construction of the hyperfine part 
of the interaction potential of particles in the case of S states. 
We use projection operators on the states of two particles with 
spin S=0 and S=1 \cite{apmplb}:
\begin{equation}
\label{eq:5}
\hat\Pi_{S=0}[u(0)\bar v(0)]_{S=0}=\frac{1+\gamma^0}{2\sqrt{2}}\gamma_5,~~~~~
\hat\Pi_{S=1}[u(0)\bar v(0)]_{S=1}=\frac{1+\gamma^0}{2\sqrt{2}}\hat\varepsilon,
\end{equation}
where $\varepsilon^\mu$ is the polarization vector of state $^3S_1$. 
The introduction (\ref{eq:5}) avoids the cumbersome multiplication of the Dirac 
bispinors and immediately proceeds to calculate the trace from the factors 
in the numerator of the interaction amplitude:
\begin{equation}
\label{eq:6}
{\cal N}=k_\alpha t_\beta\varepsilon^{\mu\nu\alpha\beta}Tr[(\hat q_1+m_l)\gamma^\nu
(\hat p_1-\hat k+m_l)\gamma^\mu(\hat p_1+m_l)\hat\Pi(\hat p_2-m_p)\gamma_5(\hat q_2-m_p)\hat\Pi^+],
\end{equation}
where $p_{1,2}$ are muon and proton four-momenta of initial state,
$q_{1,2}$ are muon and proton four-momenta of final state, $t=p_1-q_1$ is the pion
four momentum.
For the calculation and simplification (\ref{eq:6}) the Form \cite{form} package is used. 
Introducing instead of $p_{1,2}$, $q_{1,2}$ the total and relative momenta of the 
particles in the initial state $p = (0, {\bf p})$ and in the final state $q = (0,{\bf q})$, 
and also taking into account their smallness for particles in the bound state 
($|{\bf p}|\sim \mu\alpha $, $|{\bf q}|\sim\mu\alpha$), 
we retain in ${\cal N} $ only the main contribution proportional to the second 
power of the transmitted 4-momentum $t = p-q $:
\begin{equation}
\label{eq:7}
{\cal N}^{hfs}=\frac{512}{3}m_l^2m_p\left[t^2k^2-(tk)^2\right].
\end{equation}
Note that the index "hfs" denotes the selection of the hyperfine part 
in (\ref{eq:6}) using the projection operators (\ref{eq:5}).

As a result, the hyperfine part of the potential of the one-pion interaction 
of a muon and a proton in the S-state takes the form:
\begin{equation}
\label{eq:8}
\Delta V^{hfs}({\bf p},{\bf q})=\frac{\alpha^2}{6\pi^2}\frac{g_p}{m_pF_\pi}
\frac{({\bf p}-{\bf q})^2}{({\bf p}-{\bf q})^2+m_\pi^2}
{\cal A}(t^2),
\end{equation}
where
\begin{equation}
\label{eq:8a}
{\cal A}(t^2)=\frac{2i}{\pi^2t^2}\int d^4k\frac{t^2k^2-(tk)^2}{k^2(k-t)^2(k^2-2kp_1)}
F_{\pi\gamma^\ast\gamma^\ast}(k^2,(k-t)^2).
\end{equation}
The function ${\cal A}(t^2) $ is characteristic for studying the imaginary 
and real parts of the amplitude of the decay of pseudoscalar mesons into a lepton pair 
\cite{bergstrom1,bergstrom2,dorokhov2}. The dispersion relation with one subtraction for ${\cal A}(t^2)$
has the form:
\begin{equation}
\label{eq:8b}
{\cal A}(t^2)={\cal A}(0)-\frac{{\bf t}^2}{\pi}\int_0^\infty ds\frac{Im{\cal A}(s)}{s(s+{\bf t}^2)},
\end{equation}

The imaginary part of ${\cal A}(t^2)$, independent of the 
specific form of the form factor $F_{\pi\gamma^\ast\gamma^\ast}(k^2,(k-t)^2)$, 
is known (see \cite{bergstrom2} and Refs. there):
\begin{equation}
\label{eq:8c}
Im {\cal A}(t^2)
=\frac{\pi}{2\beta(t^2)}
\ln\frac{1-\beta(t^2)}{1+\beta(t^2)},
\end{equation}
where $\beta(t^2)=\sqrt{1-4m_l^2/t^2}$.

It is convenient to redefine the constant ${\cal A}(0)$ in terms of 
the moments (derivatives) of the transition form factor in the form of 
a series in the small parameter $\xi^2\equiv m_l^2/\Lambda^2$, where $\Lambda^2 $ 
is the characteristic scale of strong interactions in the transitional form factor,
\cite{Dorokhov:2008cd}
\begin{align}
& A\left( 0\right)  =\sum_{n=0}^{\infty}\frac{\left(  -\xi^{2}\right)  ^{n}%
}{n!}\frac{\Gamma_{1+2n}}{\Gamma_{1+n}\Gamma_{3+n}}\left\{  \left(
3+2n\right) \int_{0}^{\infty}dxG^{\left(n+1\right)}\left(x\right)\ln x\right.  
\label{A0a}\\
& \left.  +G^{\left(n\right)}\left(x=0\right)\left[2+\left(
3+2n\right)\left(\ln4\xi^{2}-\gamma_{E}-2\psi(n+1)+\psi(n+1/2)-\frac{2n+3}{(n+1)(n+2)}\right)\right]\right\},  \nonumber
\end{align}
where a dimensionless variable $x=k^2/\Lambda^2$ is introduced, 
$G\left(  x\right)  \equiv F_{\pi\gamma*\gamma*}
\left(k^2,k^2\right)$ и $\psi(n)$ is the digamma function.
As it was shown in \cite{dorokhov2} for the description of experimental
data on transition form factors it is sufficient to use the simplest 
monopole parametrization
\begin{equation}
G\left(  x\right)  =\frac{1}{1+x},
\label{Gt}\end{equation}
and the use of CLEO data \cite{Gronberg:1997fj} and QCD asymptotics \cite{Lepage:1980fj} 
defines the parameter $\Lambda^2 $ in the range of values
\begin{equation}
\Lambda^2=[0.448 \div 0.549]^2~GeV^2.
\label{L2}\end{equation}
With the formfactor (\ref{Gt}), the leading logarithmic contributions can be summed as
\cite{Dorokhov:2008cd}
\begin{equation}
A\left(  0\right)  = \frac{\ln\xi^2}{12\xi^4}[1+6\xi^2-\sqrt{1-4\xi^2}(1+8\xi^2)]-\frac{5}{4}+{\cal O} (\xi^2)
\label{AG}\end{equation}
Thus, for an electron, the value $A\left(0\right)$ will be equal to \cite{dorokhov2}
\begin{equation}
A\left(  0\right)  =-21.9 \pm 0.3,
\label{AGe}\end{equation}
but for a muon
\begin{equation}
A\left(  0\right)  =-6.1 \pm 0.3.
\label{AGe}\end{equation}
In the latter case, the power corrections to $\xi^6 $ should be retained 
in (\ref{A0a}), (\ref{AG}) for numerical estimates. It should also be noted that 
the effects off-shell pion are insignificant \cite{dorokhov1,Masjuan:2015cjl}. 
The maximum precise definition of the numerical value of ${\cal A}(0)$ is very important 
for achieving high accuracy of calculation.

Going then to (\ref{eq:8}) into a coordinate representation 
using the Fourier transform, we get the following single-pion exchange potential:
\begin{equation}
\label{eq:10}
\Delta V^{hfs}(r)=\frac{\alpha^2g_p}{6F_\pi m_p\pi^2}\Big\{{\cal A}(0)
\left[\delta({\bf r})-\frac{m_\pi^2}{4\pi r}e^{-m_\pi r}\right]-
\end{equation}
\begin{displaymath}
\frac{1}{\pi}\int_0^\infty\frac{ds}{s}Im{\cal A}(s)\left[\delta({\bf r})+
\frac{1}{4\pi r(s-m_\pi^2)}\left(m_\pi^4e^{-m_\pi r}-s^2e^{-\sqrt{s}r}\right)\right]\Bigr\}.
\end{displaymath}
We preserved in (\ref{eq:10}) the contributions of both terms of the function 
${\cal A}(t^2)$ from (\ref{eq:8b}), although numerically they can vary significantly.

Calculating the matrix elements with wave functions of $1S$ and $2S$ states, 
we obtain the corresponding contributions to the HFS spectrum in the form:
\begin{equation}
\label{eq:11}
\Delta E^{hfs}(1S)=\frac{\mu^3\alpha^5g_A}{6F_\pi^2\pi^3}\Biggl\{
{\cal A}(0)\frac{4W(1+\frac{W}{m_\pi})}{m_\pi(1+\frac{2W}{m_\pi})^2}
-\frac{1}{\pi}\int_0^\infty\frac{ds}{s}Im{\cal A}(s)\times
\end{equation}
\begin{displaymath}
\left[1+\frac{1}{4W^2(s-m_\pi^2)}\left(\frac{m_\pi^4}{(1+\frac{m_\pi}{2W})^2}-
\frac{s^2}{(1+\frac{\sqrt{s}}{2W})^2}\right)\right]\Biggr\}
=-0.0017~meV,
\end{displaymath}
\begin{equation}
\label{eq:12}
\Delta E^{hfs}(2S)=\frac{\mu^3\alpha^5g_A}{48F_\pi^2\pi^3}\Biggl\{
{\cal A}(0)\frac{W(8+11\frac{W}{m_\pi}+8\frac{W^2}{m_\pi^2}+2\frac{W^3}{m_\pi^3})}{2m_\pi(1+\frac{W}{m_\pi})^4}
-\frac{1}{\pi}\int_0^\infty\frac{ds}{s}Im{\cal A}(s)\times
\end{equation}
\begin{displaymath}
\left[1+\frac{1}{(s-m_\pi^2)}\left(\frac{m_\pi^2(2+\frac{W^2}{m_\pi^2})}{2(1+\frac{W}{m_\pi})^4}-
\frac{s(2+\frac{W^2}{s})}{2(1+\frac{W}{\sqrt{s}})^4}\right)\right]\Biggr\}
=-0.0002~meV,
\end{displaymath}
where the Goldberg-Treiman relation is used for the pion-nucleon interaction constant:
$g_p=g_{\pi NN}=m_pg_A/F_\pi$ with $ g_A=1.27$, $F_\pi=0.0924$ GeV, $W=\mu\alpha$.
The error in the results of \eqref{eq:11}-\eqref{eq:12} is determined by the error 
in the definition of ${\cal A}(0)$ from \eqref{AGe} and is less than 10~$\% $.
Using (\ref{eq:11})-(\ref{eq:12}), one can obtain an estimate of the contribution 
of $\eta$ mesons. These contributions, equal to (-0.0001) meV (1S), (-0.00002) meV (2S), 
yield significantly to the contribution of the pion due to the decrease in the 
interaction constant $g_{NN\eta}$. The formulas \eqref{eq:11}-\eqref{eq:12} 
can be used to estimate the corresponding contributions in the hyperfine structure 
of electron hydrogen. Thus, for the $1S$-state of the hydrogen atom, 
we obtain $ \Delta E^{hfs}(1S)=-1.25$ Hz.

The formalism of projection operators can also be used in constructing
hyperfine part of the particle interaction potential for P-states, as it was
proposed in \cite{apm2015,apm2017} (the main contribution to the hyperfine structure of the P-levels 
is given by the Breit potential in the coordinate representation (\ref{eq:3})).
We shall show this in the case of the hyperfine splitting of the $2P_{1/2} $ state,
taking into account only ${\cal A}(0)$ from (\ref{eq:8b}).
We represent the wave function of the $2P$-state in the momentum representation in the tensor form
\begin{equation}
\label{eq:13}
\psi_{2P}({\bf p})=\left(\varepsilon\cdot n_p\right)R_{21}(p),
\end{equation}
where $\varepsilon_\omega$ is the polarization vector of orbital motion, 
$n_p=(0,{\bf p}/p)$, $R_{21}(p)$ is a radial wave function in momentum representation.
Using the muon bispinor in the rest frame and the polarization vector $\varepsilon_\omega $, 
we introduce the projection operator on the muon state with the total angular momentum $ J=1/2 $:
\begin{equation}
\label{eq:14}
\hat\Pi^\omega_{\cal P}=\frac{i}{\sqrt{3}}\gamma_5(\gamma_\omega-v_\omega)\psi,
\end{equation}
where the introduced Dirac's bispinor $\psi$ describes the muon state with the total 
angular momentum $ J = 1/2 $, $v^\mu=(1,0,0,0)$.
Projecting the muon-proton pair to states with the total angular momentum $F =1,0$ 
by means of (\ref{eq:5}), we can represent the numerator of the muon-proton interaction 
amplitude (see Fig.~\ref{fig1}) as:
\begin{equation}
\label{eq:15}
{\cal N}_P=\frac{1}{3}k_\alpha t_\beta \varepsilon^{\mu\nu\alpha\beta}
Tr\Bigl[\hat\Pi (\gamma_\lambda-v_\lambda)
\gamma_5(\hat q_1+m_l)\gamma_\nu(\hat p_1-\hat k+m_l)\gamma_\mu(\hat p_1+m_l)\times
\end{equation}
\begin{displaymath}
\gamma_5(\gamma_\omega-v_\omega)\hat\Pi(\hat p_2-m_p)\gamma_5(\hat q_2-m_p)\Bigr]n_p^\omega n_q^\lambda.
\end{displaymath}
Then the potential of the hyperfine splitting of the $2P_{1/2}$ energy level can be represented 
in the momentum representation as follows:
\begin{equation}
\label{eq:16}
\Delta V^{hfs}_{2P_{1/2}}({\bf p},{\bf q})=-\frac{\alpha^2g_A}{24\pi^3F^2_\pi}\frac{({\bf p}{\bf q})
\left(\frac{p}{q}+\frac{q}{p}\right)-2pq}{({\bf p}-{\bf q})^2+m_\pi^2}{\cal A}(0).
\end{equation}
As in the previous formulas, we kept in (\ref{eq:15}) the leading contribution 
to the relative momenta ${\bf p}$, ${\bf q}$ proportional to ${\cal A}(0)$. 
The matrix element that determines the required hyperfine splitting of the $2P_{1/2}$ 
level has the form:
\begin{equation}
\label{eq:17}
\Delta E^{hfs}_{2P_{1/2}}=\int\frac{d{\bf p}}{(2\pi)^{3/2}}R_{21}(p)
\int\frac{d{\bf q}}{(2\pi)^{3/2}}R_{21}(q)\Delta V^{hfs}_{2P_{1/2}}({\bf p},{\bf q}),
\end{equation}
where the radial wave function in momentum representation has the form:
\begin{equation}
\label{eq:18}
R_{21}(p)=\frac{128}{\sqrt{3\pi}}\frac{W^{7/2}p}{(4p^2+W^2)^3}.
\end{equation}
The expression (\ref{eq:17}) contains two typical integrals that
are calculated analytically:
\begin{equation}
\label{eq:19}
I_1=\int\frac{d{\bf p}}{(2\pi)^{3/2}}R_{21}(p)
\int\frac{d{\bf q}}{(2\pi)^{3/2}}R_{21}(q)
\frac{({\bf p}{\bf q})
\left(\frac{p}{q}+\frac{q}{p}\right)}{({\bf p}-{\bf q})^2+m_\pi^2}=
\frac{2}{3}\frac{(4a+5)}{(a+2)^4},~~~a=\frac{2m_\pi}{W},
\end{equation}
\begin{equation}
\label{eq:20}
I_2=\int\frac{d{\bf p}}{(2\pi)^{3/2}}R_{21}(p)
\int\frac{d{\bf q}}{(2\pi)^{3/2}}R_{21}(q)
\frac{pq}{{\bf p}-{\bf q})^2+m_\pi^2}=
\frac{a(3a+8)+6}{2(a+2)^4}.
\end{equation}
With the help of (\ref{eq:19})-(\ref{eq:20}) we get the following analytical 
formula for splitting $2P_{1/2}$ level:
\begin{equation}
\label{eq:21}
\Delta E^{hfs}_{2P_{1/2}}=\frac{\alpha^7\mu^5g_A}{288\pi^3F_\pi^2m_\pi^2}
{\cal A}(0)
\frac{\left(9+8\frac{W}{m_\pi}+2\frac{W^2}{m_\pi^2}\right)}{(1+\frac{W}{m_\pi})^4}=
0.0004~\mu eV.
\end{equation}
The contribution of $\eta$ meson is $8\cdot 10^{-5} $ $\mu eV$. 
The numerical value of the contribution in the case of the $ 2P_ {1/2} $ 
level substantially decreases compared to the $ 2S_ {1/2} $ level, since 
the order of the contribution itself increases.
If for $2S_{1/2}$ level the order of the contribution is determined 
by the factor $\alpha^6 $, then for $ 2P_{1/2}$ level it has the 
form $\alpha^7 $. For the level $ 2P_{3/2}$, the further decrease 
in the correction value in the HFS is determined by the factor $10^2$.

\section{The positronium exchange in HFS of muonic hydrogen}

\begin{figure}[th]
\centerline{\includegraphics[scale=1.0]{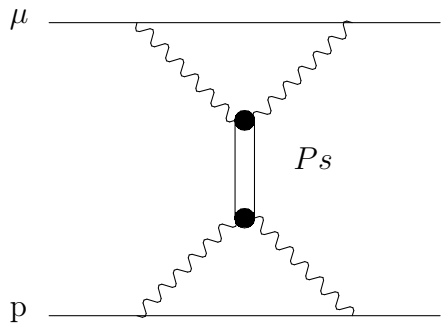}}
\caption{The amplitude of the positronium interaction in muonic hydrogen.}%
\label{fig3}%
\end{figure}

On the one hand, the single-pion exchange mechanism investigated in this paper 
gives an insignificant correction to the hyperfine splitting of the energy levels, 
which can not explain the "puzzle of the proton radius." On the other hand, 
it can be said that this correction turned out to be "unexpectedly large" in magnitude, 
referring to the exotic character of the muon-proton interaction itself. 
In this connection it was interesting to estimate the analogous contribution 
that arises as a result of the positronium exchange between a muon and a proton. 
The amplitude of such an interaction is shown in Fig.~\ref{fig3}. The estimation of the contributions 
of the hypothetical interaction with particles of mass of the order of 1 MeV both 
in the Lamb shift and in the HFS of the muonic hydrogen energy spectrum was discussed 
some time ago in \cite{barger,yavin,sgk2010} in connection with the problem of the 
proton charge radius.

\begin{figure}[th]
\centerline{\includegraphics[scale=1.]{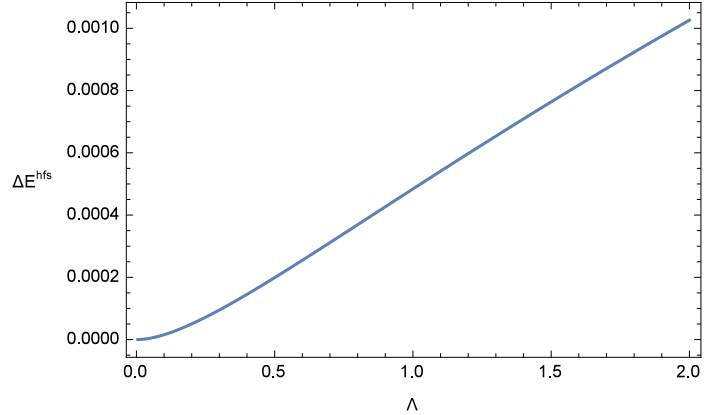}}
\caption{Hyperfine splitting of $1S$ level in meV, as a function of parameter 
$\Lambda$ in the transition form factor $Ps\to\gamma^\ast\gamma^\ast$ in GeV.}%
\label{fig4}
\end{figure}

\begin{figure}[th]
\centerline{\includegraphics[scale=1.]{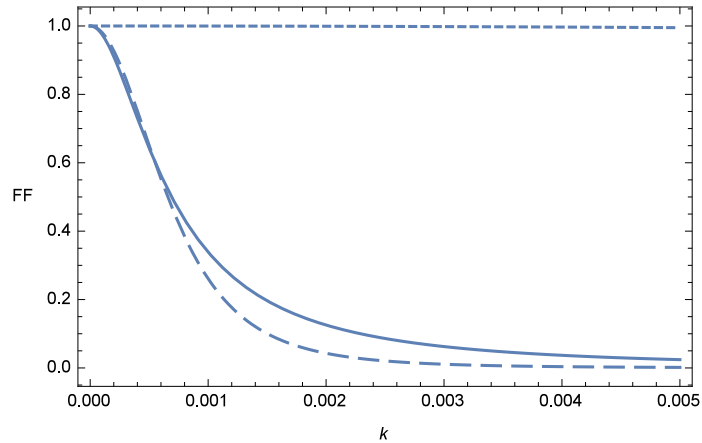}}
\caption{The transition form factor $Ps\to\gamma^\ast\gamma^\ast$, as a function of momentum
k in GeV. The solid curve denotes the perturbative form factor \eqref{eq:24}. 
The dotted curve denotes a form factor in the Vector Dominance Model with 
a cutoff parameter equal to the mass of positronium. The dotted curve 
denotes the form factor in the Vector Dominance Model with the cutoff 
parameter equal to the mass of the muon.}
\label{fig5}
\end{figure}
The potential of single-positronium exchange in muonic hydrogen for the hyperfine 
splitting of S-states in the momentum representation has the form:
\begin{equation}
\label{eq:22}
\Delta V^{hfs}_{\mu p}({\bf t})=\frac{2\alpha^2}{3\pi^2}F^2_{Ps\gamma^\ast\gamma^\ast}(0)
{\cal A}_\mu(0){\cal A}_p(0)\frac{{\bf t}^2}{{\bf t}^2+m_{Ps}^2}({\bf s}_1{\bf s}_2),
\end{equation}
where for simplicity we use the approximation  ${\cal A}_{\mu,p}({\bf t}^2)\approx{\cal A}_{\mu,p}(0)$
for the effective constants of the muon-proton interaction 
with positronium. Estimating the parameter $F_{Ps\gamma^\ast\gamma^\ast}(0)$ 
using the decay width of the positronium into two photons by the formula
\begin{equation}
\label{eq:4a}
F_{Ps\gamma^\ast\gamma^\ast}(0)=\sqrt{\frac{64\pi\Gamma(Ps\to\gamma\gamma)}
{(4\pi\alpha)^2m_{Ps}^3}},
\end{equation}
where $\Gamma(Ps\to\gamma\gamma)$ is the width of the positronium decay into a pair of photons, 
we find the contribution of this interaction to the hyperfine structure in the form:
\begin{equation}
\label{eq:23}
\Delta E^{hfs}_{Ps}(1S)=\frac{\mu^3\alpha^8}{6\pi^4m_e^2}{\cal A}_{\mu}(0){\cal A}_{p}(0)
\frac{(1+\frac{m_{Ps}}{W})}{\left(1+\frac{m_{Ps}}{2W}\right)^2}.
\end{equation}
Using further the expression in the Vector Dominance Model for ${\cal A}_{\mu, p}(0) $ 
(we introduce the dimensionless loop momentum using the parameter $\Lambda$) 
and calculating the integral with the Feynpar package \cite{west},
\begin{equation}
\label{eq:9}
{\cal A}_{\mu,p}(0)=\int\frac{6id^4k}{(4\pi^2)}\frac{1}{(k^2)^2(k^2-1)^2(k^2-2k\tilde p_{1,2})}=
-\frac{3}{2}\frac{\ln\Bigl[\frac{1-2\frac{m_{l,p}^2}{\Lambda^2}+
\sqrt{1-\frac{4m_{l,p}^2}{\Lambda^2}}}{2\frac{m_{l,p}^2}{\Lambda^2}}
\Bigr]}{\sqrt{1-\frac{4m_{l,p}^2}{\Lambda^2}}}.
\end{equation}
we obtain the numerical values of the contributions to the hyperfine structure. 
It is convenient to represent the result of the calculation of $\Delta E^{hfs}_{Ps}(1S)$ 
on the graph as a function of the cutoff parameter $\Lambda$ (see Fig.~\ref{fig4}). 
Summation over various excited states of the positronium
gives an additional factor $\sum_0^\infty 1/n^3=1.202$. In the perturbative 
loop theoretical model, the form factor of the transition of two photons to the positronium 
is determined by the following tensor integral
\begin{equation}
\label{eq:loop}
I^{\mu\nu}=\int\frac{d^4q}{(2\pi)^4}\frac{Sp[\gamma_5(\hat q+\hat k+m_e)\gamma^\mu
(\hat q+m_e)\gamma^\nu(\hat q-\hat t+\hat k+m_e)]}
{(q^2-m_e^2)[(q+k)^2-m_e^2][(q-t+k)^2-m_e^2]}.
\end{equation}
Using the Feynman parametrization in calculating the loop integral and 
setting $t=0$ in \eqref{eq:loop}, we obtain the following expression 
for the transition form factor:
\begin{equation}
\label{eq:24}
F_{Ps\gamma^\ast\gamma^\ast}(k^2,k^2)=\frac{\alpha^{3/2}}{m_e\sqrt{\pi}}
\frac{1}{k^2}\Biggl[-Li_2(\frac{2k}{\sqrt{k^2-4}-k})-Li_2(-\frac{2k}{\sqrt{k^2-4}+k})+
\end{equation}
\begin{displaymath}
Li_2(\frac{2k}{k-\sqrt{k^2+4}})+Li_2(\frac{2k}{\sqrt{k^2+4}+k})\Biggr],
\end{displaymath}
where the dimensionlessness of the integral is carried out with the help of the 
electron mass $m_e$.
If we compare \eqref{eq:24} and the transition factor in the Vector
Dominance Model, it can be noted that the mass of positronium acts as a natural 
cutoff parameter. Such a form factor decreases rapidly with increasing virtuality $k^2$ 
and the magnitude of the correction $\Delta E^{hfs}_{Ps}(1S)$ is negligible. 
As the cutoff parameter grows, the contribution increases logarithmically and starting 
with $\Lambda\sim 1 $ GeV can already have such a value, which must be taken into account for
more accurate determination of the total hyperfine splitting. An increase in the value 
of the cutoff parameter in the transition form factor means that the 
positronium production probability for large photon virtualities $k^2$ and $(t-k)^2$ 
remains significant.

\section{Conclusion}

The high precision measurement of the hyperfine splitting of the muonic
hydrogen atom ground state is planned in near future (see,
\cite{tomalak,pohl_2017,ma_2017,adamczak_2017}). The experiment of 
FAMU (Fisica Atomi MUonici) collaboration \cite{adamczak_2017} aims to investigate of the proton
radius puzzle and determination of the Zemach radius with HFS 
of $(\mu^- p)_{1S}$ and to achieve unprecedented accuracy
$\delta\lambda/\lambda\le 10^{-5}$. Even higher experimental resolution for the
$\Delta E_{exp}^{hfs}$ 2 ppm is expected to obtain in \cite{ma_2017}.
Taking into account that the value of the ground state hyperfine splitting in 
muonic hydrogen is equal 182.725 meV \cite{apm2005} (see also \cite{pineda2017})
the planned increase in the accuracy of measuring the hyperfine structure 
of the spectrum in muonic hydrogen will make it possible to verify various 
theoretical contributions of higher order, and, possibly, to reveal new terms in the particle 
interaction operator.

In this paper, we investigate the contribution of a pseudoscalar meson  
to the potential of the hyperfine interaction of the muon and the proton 
and into the hyperfine structure of the energy spectrum. In the framework 
of the quasipotential method in quantum electrodynamics and the use of the 
technique of projection operators on the states of two particles with 
a definite spin, we constructed particle interaction operators
\eqref{eq:10}, \eqref{eq:16} and obtained analytical expressions 
for the hyperfine splittings of the S and P energy levels 
\eqref{eq:11}, \eqref{eq:12}, \eqref{eq:21}.
Numerical estimates of the contributions \eqref{eq:11}, \eqref{eq:12}, 
\eqref{eq:21} connected with the exchange of pseudoscalar mesons are made 
on their basis. An important role in the numerical calculation of the studied 
contributions is played by the function ${\cal A}(t^2)$ \eqref{eq:8b} 
related with the form factor of the transition of two photons to 
a pseudoscalar meson \eqref{eq:4}.
For more accurate determination of the constant ${\cal A}(0)$ in 
\eqref{eq:8b}, we used the results of the works 
\cite{dorokhov1,dorokhov3,Dorokhov:2009xs} in which $ {\cal A}(0)$ 
is defined in terms of the moments of the transition form factor.
We also obtained numerical estimates of the contribution \eqref{eq:23} 
to the hyperfine structure of the spectrum due to positronium exchange.

The obtained analytical results are in agreement with the previous 
calculations of this effect in the framework of chiral perturbation theory 
\cite{pascalutsa,kou,pang}. The numerical result for the hyperfine splitting of the 
2S state $(-0.09 \pm 0.06)$~$\mu eV$ from \cite{kou}
is comparable to our value (-0.0002) meV, taking into account the theoretical error, 
and our result for HFS $2P_{1/2} $ practically coincides with the value $3.7\cdot 10^{-4}$ $\mu eV$ from 
\cite{kou}. The difference from the result of \cite{kou} for 2S-level is due 
to taking into account in \cite{kou} the dependence of the vertex function of the 
pion-nucleon interaction on the transmitted momentum.

Using the obtained result for the hyperfine interaction of a muon and a proton due 
to a one-pion exchange, it is possible to estimate the same contribution in the case 
of other light muonic atoms, for example
muonic deuterium. The simplest approximation in describing the pion-deuteron 
interaction is that the deuteron is regarded as a state of two almost free
nucleons, and the spins of the neutron and proton in the sum give the total spin S=1 
of the deuteron. Consequently, it can be concluded that the contribution 
of the pion-neutron interaction to the hyperfine structure of muonic deuterium 
is the same as that of the pion-proton one, and the total contribution 
to the hyperfine splitting, for example, of the 2S level, is twice that, that is,
has a value of (-0.0004) meV.

\begin{acknowledgments}
We are grateful to O.~Tomalak for useful communication.
The work is supported by Russian Science Foundation
(grant No. RSF 15-12-10009) (A.E.D.), 
the Chinese Academy of Sciences visiting professorship for senior international scientists 
(grant No. 2013T2J0011) (N.I.K.), Russian Foundation for Basic Research
(grant No. 16-02-00554) (A.P.M., F.A.M.)
\end{acknowledgments}

\end{document}